# An ultra-sensitive pulsed balanced homodyne detector: Application to time-domain quantum measurements

H. Hansen, T. Aichele, C. Hettich, P. Lodahl, A. I. Lvovsky, J. Mlynek and S. Schiller*

*Fachbereich Physik, Universität Konstanz, M696, D-78457 Konstanz, Germany*

A pulsed balanced homodyne detector has been developed for precise measurements of electric field quadratures of pulsed optical quantum states. A high level of common mode suppression (> 85 dB) and low electronic noise (730 electrons per pulse) provide a signal to noise ratio of 14 dB for the measurement of the quantum noise of individual pulses. Measurements at repetition rates up to 1 MHz are possible. As a test, quantum tomography of the coherent state is performed and the Wigner function and the density matrix are reconstructed with a 99.5% fidelity. The detection system can also be used for ultrasensitive balanced detection in cw mode, e.g. for weak absorption measurements.

The rapidly developing field of quantum information technology requires reliable means of characterizing optical quantum states. In application to nonclassical light, balanced homodyne detection (BHD) has proved invaluable for the direct measurement of electric field quadratures of electromagnetic modes. Proposed in 1983 by Yuen and Chan[1], BHD was initially used to detect squeezed states of the electromagnetic field[2], and later for studies such as complete characterization of quantum states via quantum tomography[3,4,5], establishing Einstein-Podolsky-Rosen type quantum correlations[6], continuous-variable quantum teleportation[7]. Very recently, BHD has been employed to demonstrate non-classical properties of electromagnetic fields in cavity QED[8]. BHD is expected to play a major role in quantum information processing in the future.

To date, most BHD measurements have been performed in the frequency domain. A significant drawback of this approach is that it reveals the information about the quantum state only within the sideband chosen for the measurement. This makes the method incompatible with other techniques of quantum state characterization (e.g. photon counting) where such precise selection of spectral modes is impossible.

Time domain BHD resolves this limitation. It was first employed by Raymer and co-workers in their pioneering experiments on quantum tomography and quantum correlations[3,9]. Their homodyne detector could resolve shot (vacuum) noise of individual pulses with a signal-to-noise ratio (SNR) of 6 dB at a sub-kHz repetiton rate. The limitations arose from the electronic subtraction efficiency (LO powers limited to $4\times10^6$ photons per pulse), electronic detection noise (580 electrons per pulse) and the slow amplification electronics.

In this paper, we present a time domain BHD system whose characteristics substantially exceed those outlined above. We have achieved a SNR of 14 dB at a pulse repetition rate of up to 1 MHz, enabling high-accuracy quantum measurements to be performed in short time. The detector exhibits 91% quantum efficiency (compared with 85% by Raymer *et al.*).

To perform BHD, the electromagnetic wave whose quantum state is to be measured is overlapped on a beamsplitter with a relatively strong local oscillator (LO) wave in the matching optical mode. The two fields emerging from the beamsplitter are incident on two high-efficiency photodiodes whose output photocurrents are subtracted. The photocurrent difference is proportional to the value of the electric field operator $\hat{E}_\theta$ in the signal mode, $\theta$ being the relative optical phase of the signal and LO.

In traditional, frequency-domain BHD a certain frequency component (typically around 5-10 MHz) of the difference signal is used to determine the quadrature quantum noise of the optical state. The measurement frequency is normally chosen so as to optimize the ratio of quantum optical noise and the electronic noise floor of the BHD. The working range is restricted from below by the $1/f$ technical noises and from above by the bandwidth of the detector electronics. Frequency domain BHD has been successfully applied both in the continuous-wave[1,2,4,10] and pulsed[11] regime with a typical SNR of 20 dB. This allowed to measure squeezed optical states with quantum noise reduction up to 7 dB[10]. When applied to pulsed sources, the frequency-domain BHD technique implies that averaging over many individual laser pulses takes place.

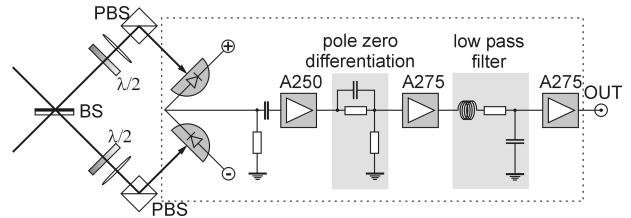

Fig. 1. Electrooptical scheme of the homodyne system

In time-domain BHD each laser pulse generates a signal which is observed in real time and yields a single value of a field quadrature. Repeated measurements on a large number of laser pulses will yield a quantum probability distribution associated with the quadrature. When transform-limited LO pulses are used, time-domain BHD will yield the *complete* information about the quantum state in the spatio-temporal optical mode matching that of the local oscillator. Time domain BHD is however more technically challenging than its frequency domain counterpart. On one hand, the electronics must ensure time resolution of individual laser pulses. On the other hand, the measured quadrature values must not be influenced by low frequency noises. The detector must thus provide ultra-low noise, high subtraction and flat amplification profile in the *entire frequency range* from

DC to at least the LO pulse repetition rate.

The electrooptical scheme of our detector is shown in Fig. 1. We used 1.6-ps, 790-nm pulses from a Spectra Physics Tsunami laser as the local oscillator. The laser was employed in combination with a pulse picker which reduced the repetition rate to 200-800 kHz. The orientation of the beamsplitter slightly deviated from 45° to provide a splitting ratio closest to 50%. The two beamsplitter outputs passed through a pair of $\lambda/2$ plates and polarizing beamsplitter (PBS) cubes, which in combination served as variable attenuators used to compensate for slightly different quantum efficiencies of the photodiodes. The two beams were then focused on a pair of Hamamatsu S3883 Si-PIN photodiodes of 300 MHz bandwidth and 94% quantum efficiency which were chosen because of their low noise equivalent power of $6.7\times10^{-15}$ W·Hz$^{-1/2}$. The polarizing beamsplitter cubes, lenses and $\lambda/2$ plates were AR coated for 790 nm so that the total optical losses did not exceed 4%.

The positive and negative charges produced by the optical pulses were collected and physically subtracted at a 470-pF capacitance which is much larger than the capacitances of the photodiodes (6 pF). The difference charge was then pre-amplified using a 2SK152 FET in connection with the low-noise Amptek A250 preamplifier and further amplified using a 5-pole pulse shaping amplifier based on two low-noise Amptek A275 amplifiers. The entire detector electronics, including the photodiodes, was built on a single PC board inside a metal box 1"×2"×4". The two photodiodes were mounted at a distance of only 1 cm from each other to minimize spurious RF interferences.

Fig. 2(a) shows typical time traces of the homodyne detector difference signal for vacuum signal input and an average LO power of $1.6\times10^8$ photons per pulse. The width of a single electrical pulse is less than 1 μs. Its peak value is a single measurement of the electric field quadrature of the signal wave. The (shot) noise of the optical pulses is highly visible on top of the electronic noise background. It indicates the statistical distribution of the quadratures of the vacuum field. For the local oscillator in a coherent state and vacuum in the signal input the statistical distribution is Gaussian.

To prove that the pulsed noise shown in Fig. 2(a) is indeed shot noise we have performed a number of tests to make sure that a) the output rms noise scaled as square root of the LO power, b) the noise observed away from multiples of the repetition rate was frequency independent (white) and c) the observed noise power coincided well with the expected magnitude.

Fig. 2(b) shows the standard deviation of the pulsed noise as a function of the local oscillator power. After subtraction of the noise background corresponding to 730 electrons/pulse, the standard deviation of the noise scales with the square root of the local oscillator power as predicted for shot noise. Such behavior was observed for local oscillator powers spanning over more than two orders of magnitude, up to $3\times10^8$ photons per pulse. This corresponds to a maximum subtraction of 85 dB.

The detector amplifier gain and linearity have been verified by connecting a 1 pF capacitor to the preamplifier input and inserting small controlled amounts of charges by applying voltage steps of a known size to the capacitor. The detector electronics exhibited excellent linearity within the required dynamic range. The measured gain value allowed to check the absolute magnitude of the observed pulsed noise. It was found to correspond to the expected shot noise magnitude to within the precision of the charge insertion capacitor.

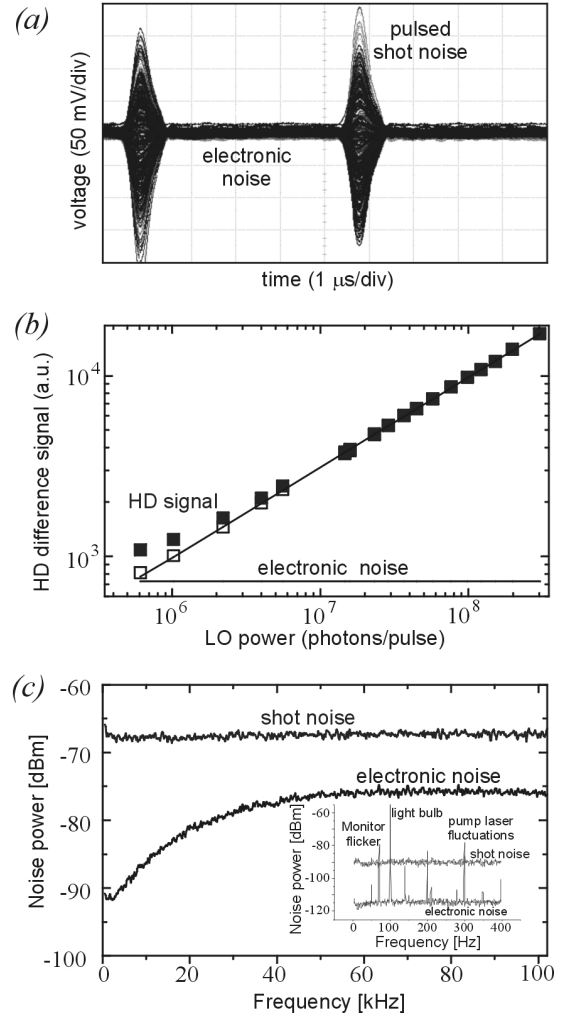

Fig. 2 (a) Oscilloscope trace of the homodyne detector output obtained at a laser repetition rate of 204 kHz and the local oscillator power of $1.6\times10^8$ photons per pulse. Each laser pulse produces a time-resolved quantum noise sample. (b) RMS peak amplitude of the noise pulses as a function of local oscillator power showing the expected square root power dependence up to local oscillator intensities of $3\times10^8$ photons per pulse. The filled squares show the measured noise variances, the open squares have been obtained from these values by subtracting the noise background. (c) Frequency-resolved noises at a LO power of $2.3\times10^7$ photons per pulse. Inset: the low frequency spectrum illustrates the effect of background light sources.

The frequency spectrum of the detector output is shown in Fig. 2(c) and is to a very good approximation white. These measurements show that the detector system could also be fruitfully used in a completely different regime, namely to perform shot-noise limited measurements with cw light at subnanowatt power levels. The inset in Fig.2

(c) shows the extreme sensitivity of the detector to external light sources.

As an application of the BHD system we performed quantum tomography of a coherent state and reconstructed its Wigner function and density matrix. To this end we used pulses from the source laser which were mode matched with the local oscillator and attenuated almost to the single-photon level. We employed the pulse picker synchronization signal to trigger the data acquisition and used an additional piezo-mounted mirror in the seed beam path to scan the relative phase $\theta$ between the seed and LO beam.

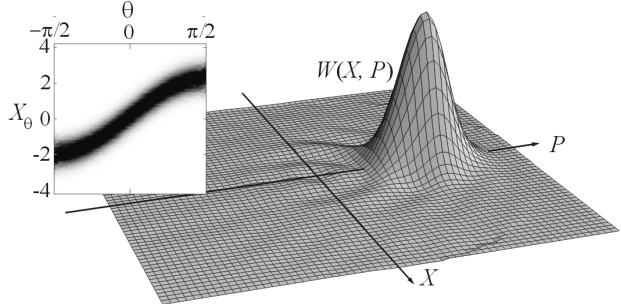

Fig. 3. Raw quadrature data and the reconstructed Wigner function of a coherent state with the excitation $\alpha = 2.24$

In an experimental run, 262144 experimental points were measured to provide 64 marginal distributions with 128 bins each (Fig. 3, inset). A 204-kHz pulse repetition rate was used, so the acquisition of each marginal distribution took about 20 ms. Since the setup was not interferometrically stable (we measured 8° average phase drift over time intervals corresponding to the complete measurement), in each distribution, the actual value of $\theta$ during the measurement was calculated from the average of all quadrature values in the distribution.

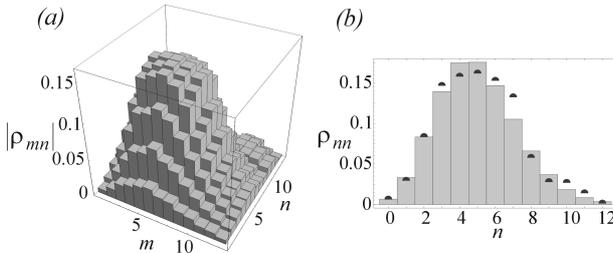

Fig. 4. (a) Absolute value of the density matrix elements in the Fock basis obtained by quantum state sampling of the marginal data shown in Fig. 3; (b) the photon number distribution. A Poissonian distribution with the same average $n$ is shown for comparison (black dots).

We have applied the inverse Radon transformation[12] with a cutoff frequency of 7.25 to the marginal data to obtain the Wigner function of the quantum state measured. The reconstructed Wigner function reflects the expected 2-dimensional Gaussian phase space distribution with a width equal to that of the vacuum state Wigner function. The ripples at the side of the main peak are numerical artifacts and arise due to various measurement errors.

The quantum sampling method[13] was applied directly to the raw experimental data to reconstruct the density matrix $\rho_{mn}$ of the state (Fig. 4). An average photon number of $\langle n \rangle = \sum n\rho_{nn} = 5.01$ photons per pulse corresponding to a coherent excitation of $\alpha = 2.24$ is inferred from the measured photon number distribution. In contrast to tomography in the cw regime,[4] these numbers do not refer to a frequency side band of a bright coherent beam but directly correspond to an average number of photons in the signal laser pulse. The comparison of the reconstructed density matrix with that of an ideal coherent state with an amplitude $\alpha = 2.24$ yields a state preparation fidelity of $F = \langle \alpha | \hat{\rho} | \alpha \rangle = 0.995$.

The BHD system was also used for an experiment on quantum tomography of the single-photon Fock states. The optical pulses containing single photons were prepared using conditional measurements on photon pairs that are born in the process of parametric down-conversion. The phase-averaged Wigner function reconstructed in the measurement showed a strong dip reaching classically impossible negative values. Further details on this experiment will be published elsewhere[14].

In conclusion, we have designed and built a pulsed balanced homodyne detector for highly accurate time-domain quantum measurements. It exhibits a high bandwidth (1 MHz), > 90% quantum efficiency, and very large (>85 dB) common mode rejection. As a demonstration of the capability of our measurement system, we have shown quantum tomography of a pulsed coherent state with a total measurement time of only 1.3 s. This detector will be of significant use in a variety of experiments in quantum optics. The detector could also be adapted for wide-band conventional balanced detection (separate beams of equal strength on each photodiode) in the frequency domain, thanks to the very low spectral noise.

We are grateful to S. Eggert for technical assistance. A. L. is supported by the A. von Humboldt Foundation.
*Present address: Institut für Experimentalphysik, Universität Düsseldorf, D-40225 Düsseldorf, Germany